\documentclass[12pt]{article}

\usepackage{amsmath,amssymb}
\usepackage{graphicx}
\usepackage{siunitx}
\usepackage{authblk}
\usepackage[margin=2.5cm]{geometry}
\providecommand{\keywords}[1]
{
  \small	
  \textbf{\textit{Keywords---}} #1
}
\usepackage[superscript]{cite}

\title{Metasurface Engineering with Tantalum Pentoxide-Coated Microspheres:\\
	Tailoring Optical Resonances and Enhancing Local Density of States}

\author[1]{Alexandra~F\u{a}l\u{a}ma\c{s}}
\author[1]{Ana~Maria~Mihaela~Gherman}
\author[2]{Renaud~Vall\'ee*}
\author[1]{Cosmin~Farc\u{a}u*}

\affil[1]{National Institute for Research and Development of Isotopic and Molecular Technologies, 65--103 Donath, 400293 Cluj-Napoca, Romania, email: cfarcau@itim-cj.ro}
\affil[2]{CNRS, University of Bordeaux, CRPP-UMR 5031, 33600 Pessac, France, email: renaud.vallee@itim-cj.ro}

\date{}

\begin{document}
	\maketitle
	
	\begin{abstract}
		Hexagonally close-packed polystyrene (PS) microsphere monolayers coated with tantalum pentoxide (Ta$_2$O$_5$) form scalable dielectric metasurfaces that support tunable photonic resonances and enhanced local density of optical states (LDOS). Here we combine fabrication, optical and fluorescence spectroscopy, and multi-scale electromagnetic simulations to quantify how the thickness of a Ta$_2$O$_5$ shell controls far-field resonances and Rhodamine 6G (Rh6G) emission.
		
		Experimentally, Ta$_2$O$_5$ shells of \SIrange{10}{70}{nm} deposited on PS microsphere lattices generate resonances that shift red with the thickness of the shell and systematically enhance the Rh6G fluorescence relative to flat Ta$_2$O$_5$ films. The largest integrated enhancement is obtained for \SIrange{30}{50}{nm} shells, where lattice resonances overlap the Rh6G excitation and emission bands.
		
		Finite-cluster finite-difference time-domain simulations reproduce the measured transmittance and reflectance spectra, confirming that the fabricated lattices are well described by the geometry of Ta$_2$O$_5$ shells covering the sphere lattice. Periodic-cell simulations of single electric dipoles yield wavelength-dependent Purcell factors $F_{\mathrm{p}}(\lambda)$ and directional $\beta$-factors $\beta_{\mathrm{top}}(\lambda)$, from which we construct emission-weighted figures of merit that link LDOS modulation to the experimentally accessible top-side fluorescence enhancement.
		
		As a complementary test of our emitter--environment model, we compare simulated and measured Purcell factors for PS/Ta$_2$O$_5$ microsphere lattices with the same Ta$_2$O$_5$ thicknesses. A physically motivated averaging scheme that accounts for emitter position, orientation and ensemble spectral smoothing yields very good agreement across all shells. Overall, our results establish Ta$_2$O$_5$-coated microsphere lattices as robust dielectric substrates for surface-enhanced fluorescence and clarify how shell thickness and emitter placement jointly control photonic resonances, LDOS and fluorescence response.
	\end{abstract}

    
	\section{Introduction}
	
	Dielectric metasurfaces based on high-index nanostructures have emerged as versatile platforms for controlling light at the wavelength scale.\cite{Kuznetsov2016,Staude2017,Liu2016} By tailoring the geometry and refractive index of subwavelength features, these structures can shape phase, amplitude and polarization and can strongly modify the local density of optical states (LDOS) seen by nearby emitters.\cite{NovotnyHecht,LodahlRMP} In contrast to plasmonic systems, high-index dielectrics provide strong field confinement with negligible Ohmic loss, which is particularly attractive for enhanced light emission and integrated photonics.\cite{Lakowicz,Kuehn2006}

    Surface-enhanced fluorescence (SEF) has been pursued primarily through plasmonic substrates such as metal nanoparticles, nanoapertures, and metal-film-over-nanosphere geometries, where strong near-field enhancement boosts excitation rates and modifies LDOS. However, the proximity of emitters to metal surfaces introduces non-radiative quenching that fundamentally limits performance for high quantum-yield dyes, and spectral tunability typically requires redesigning the nanostructure geometry. Dielectric alternatives based on photonic crystals or Mie-resonant silicon structures can modulate LDOS without quenching losses, but are generally confined to the near-infrared or require costly lithographic patterning incompatible with large-area fabrication.\cite{Kuznetsov2016,Staude2017}
	
	For a dipole emitter coupled to a structured photonic environment, the spontaneous-emission rate $\Gamma(\lambda)$ at wavelength $\lambda$ is proportional to the projected LDOS. The Purcell factor is defined by Eq.~\eqref{eq:Purcell}, 
	\begin{equation}
		F_{\mathrm{p}}(\lambda) = \frac{\Gamma(\lambda)}{\Gamma_0(\lambda)} =
		\frac{P(\lambda)}{P_0(\lambda)},
		\label{eq:Purcell}
	\end{equation}
	where $\Gamma_0$ and $P_0$ are the decay rate and radiated power of a reference dipole in a homogeneous medium.\cite{NovotnyHecht,LodahlRMP} The fraction of the emitted power directed into a specific detection channel, such as the upper half-space above a metasurface, is quantified by a directional $\beta$-factor (Eq.~\eqref{eq:beta})
	\begin{equation}
		\beta_{\mathrm{top}}(\lambda) = \frac{P_{\mathrm{top}}(\lambda)}{P(\lambda)},
		\label{eq:beta}
	\end{equation}
	where $P_{\mathrm{top}}$ is the power radiated into the superstrate.
	
	For a high-quantum-yield dye such as Rhodamine 6G (Rh6G) in a low-loss dielectric environment, non-radiative quenching is weak and the detected emission spectrum above a metasurface can be approximated as\cite{Lakowicz,LodahlNature2004}
	\begin{equation}
		I_{\mathrm{det}}(\lambda) \propto S(\lambda)\, F_{\mathrm{p}}(\lambda)\, \beta_{\mathrm{top}}(\lambda),
		\label{eq:Idet}
	\end{equation}
	where $S(\lambda)$ is the intrinsic emission spectrum of the dye, normalized to unit area. Equation~\eqref{eq:Idet} highlights that both radiative-rate enhancement and directional out-coupling determine the observed emission.
	
	For broadband emitters, it is convenient to define emission-weighted figures of merit (Eqs.~\eqref{eq:Fpem} and~\eqref{eq:Fpemrad})
	\begin{align}
		F_{\mathrm{p}}^{\mathrm{em}} &= \int S(\lambda)\, F_{\mathrm{p}}(\lambda)\, \mathrm{d}\lambda, \label{eq:Fpem}\\
		F_{\mathrm{p,rad}}^{\mathrm{em}} &= \int S(\lambda)\, F_{\mathrm{p}}(\lambda)\, \beta_{\mathrm{top}}(\lambda)\, \mathrm{d}\lambda.
		\label{eq:Fpemrad}
	\end{align}
	For a high-yield emitter, the overall top-side emission enhancement factor is then given approximately by Eq.~\eqref{eq:EFtop},
	\begin{equation}
		E_{\mathrm{F,top}} \simeq F_{\mathrm{p,rad}}^{\mathrm{em}}.
		\label{eq:EFtop}
	\end{equation}
	These quantities provide a direct bridge between far-field resonances observable in reflection and transmission, the near-field LDOS, and the measured fluorescence.\cite{LodahlRMP}
	
	Self-assembled monolayers of PS microspheres form hexagonally close-packed colloidal crystals over large areas at low cost.\cite{Mie1908} Coated with metal overlayers, these arrays are well-established SEF and SERS platforms, whose plasmonic resonances are tunable through the metal thickness and type.\cite{Nechita,Kuehn2006} Replacing the metal with a high-index dielectric overlayer could transform these systems into photonic crystals metasurfaces that support Mie-type and guided-mode resonances, but this direction has remained largely unexplored.\cite{Kuznetsov2016,Staude2017} To the best of our knowledge, the combination of colloidal self-assembly with a high-index dielectric coating to produce a functional metasurface has not been explored. A key challenge is that, unlike plasmonic systems where richer empirical and theoretical toolkits exist, dielectric colloidal-based metasurfaces currently lack a quantitative design framework that connects fabrication geometry to LDOS and measurable fluorescence response.
    
    Tantalum pentoxide (Ta$_2$O$_5$, tantala) is a strong candidate for this role. It combines a refractive index around 2.0--2.2 across the visible spectrum with near-zero absorption, a small thermo-optic coefficient, and full compatibility with standard thin-film deposition and CMOS fabrication.  \cite{Steinlechner2018,Belt2017,Wang2021,Subramanian2009,ZhangLSA} Ta$_2$O$_5$ is widely used in high-reflectivity coatings for gravitational-wave detectors,\cite{Steinlechner2018} in low-loss waveguides and microresonators,\cite{Belt2017,Wang2021,Subramanian2009} and more recently in high-performance metasurfaces for focusing, holography and structural colour.\cite{ZhangLSA} Ta$_2$O$_5$ waveguides have also been exploited for fluorescence detection in evanescent-field configurations.\cite{Schmitt2008,Grandin2006} In this context, Ta$_2$O$_5$ offers a low-loss alternative to metals for SEF, capable of enhancing local fields and LDOS without the strong non-radiative quenching typical of plasmonic substrates.\cite{Lakowicz,Kuehn2006} Compared to other high-index dielectrics, Ta$_2$O$_5$ combines a higher refractive index than Si3N4 with broader transparency than TiO$_2$, and lower deposition complexity than HfO$_2$, making it particularly well suited for conformal visible-range coatings on polymer microsphere lattices. Its use in colloidal-based photonics for LDOS engineering, however, has not been reported.

	In this work, we address this gap by establishing a complete and experimentally validated design framework for Ta$_2$O$_5$-coated PS microsphere lattices as dielectric SEF metasurfaces. We show that the Ta$_2$O$_5$ shell thickness alone, a single geometric parameter controllable through deposition time, continuously tunes the lattice photonic resonance across approximately \SI{80}{\nano\meter} of the visible spectrum, enabling on-demand alignment with a chosen emitter band. We construct a multi-scale physical framework that links this geometric parameter to ensemble-measurable quantities through three connected layers: finite-cluster finite-difference time-domain (FDTD) simulations that reproduce experimental far-field spectra and implicitly validate the conformal shell geometry; periodic-cell dipole simulations that separately quantify the Purcell factor $F_{\mathrm{p}}(\lambda)$ and the directional  $\beta$-factor $\beta_{\mathrm{top}}(\lambda)$; and a physically motivated emitter-placement averaging model, grounded in capillary wetting arguments, that translates single-emitter LDOS calculations into ensemble Purcell factors in quantitative agreement with time-resolved fluorescence measurements. The decoupling of steady-state and time-resolved fluorescence responses across shell thicknesses further reveals the distinct roles of excitation enhancement, LDOS modification, and emission out-coupling efficiency, factors that are independently addressable through geometry in this platform. Together, these results establish not only Ta$_2$O$_5$-coated microsphere lattices as a low-loss, scalable SEF substrate, but more broadly a transferable design approach for dielectric colloidal metasurfaces that can be extended to other high-index materials and emitter systems without the need for iterative experimental optimization.

	\section{Experimental and numerical methods}
	
	\subsection{Fabrication of Ta$_2$O$_5$-coated microsphere lattices}
	
	Hexagonally close-packed PS microsphere monolayers were prepared on glass substrates following standard colloidal self-assembly protocols.\cite{Zhang} Monodisperse PS spheres of nominal diameters $D = \SI{460}{nm}$ were used. After formation of the monolayer, Ta$_2$O$_5$ was deposited by e-beam evaporation (Kenosistec KE400). The nominal Ta$_2$O$_5$ thicknesses were $t_\mathrm{shell} = \SI{10}{nm}, \SI{30}{nm}, \SI{50}{nm}$ and \SI{70}{nm}, as monitored by quarts crystal microbalance during deposition.
	
	These coatings produce hybrid PS/Ta$_2$O$_5$ shells whose effective optical thickness increases with $t_\mathrm{shell}$. The same thicknesses were used in the simulations described below.
	
	\subsection{Optical characterization}
	
	Normal-incidence transmission and reflection spectra in the visible range were measured with a UV--vis spectrophotometer equipped with focusing optics adapted to the finite size of the microsphere patches. Transmission $T(\lambda)$ was obtained by normalizing the transmitted intensity to that of a glass reference. Reflectivity $R(\lambda)$ was measured in epi configuration and normalized to a suitable standard.
	
	\subsection{Fluorescence measurements}
	Rhodamine~6G (Rh6G) was dissolved in a 4\% w/w polyvinylpyrrolidone (PVP) solution in ethanol and spin-coated onto the Ta$_2$O$_5$ samples and a glass reference, forming thin and nearly uniform fluorescent layers.\cite{Lakowicz} Fluorescence spectra were recorded using a micro-Raman spectrometer (B\&W Tek) coupled to a microscope with a 20$\times$ objective with excitation at \SI{532}{\nano\metre}. At least ten spectra were acquired at different positions from each sample, using an integration time of \SI{1}{\second} and five accumulations under low laser power to avoid photo-bleaching.

    \subsection{Time-resolved fluorescence measurements}
    Time-correlated single photon counting (TCSPC) measurements were performed on a time-resolved fluorescence set-up (Chimera, LightConversion, Lithuania) by exciting the samples at \SI{545}{\nano\meter} using a 170 fs pulsed laser working at 80 kHz pulse repetition frequency (Pharos, LightConversion, Lithuania). Emission was collected in the \SIrange{570}{630}{nm} spectral range through a 10$\times$ objective. Data were analyzed in EasyTau2 (PicoQuant) using reconvolution fitting with the measured instrumental response function (IRF); control samples were fit with mono-exponential functions and Ta$_2$O$_5$ metasurface samples with bi-exponential functions. Intensity-weighted average lifetimes were calculated using Eq.~\eqref{eq:TCSPC}: 
       \begin{equation}
		\tau_{\mathrm{avg\,Int}}=\frac{\sum_{\substack{i=1 \\ J_i >0}}^{n_{\mathrm{exp}}} I_i \tau_i}{\sum_{\substack{i=0 \\J_i >0}}^ {n_{\mathrm{exp}}} I_i}.
		\label{eq:TCSPC}
	\end{equation}

	\subsection{Electromagnetic simulations}
	
	Two complementary numerical approaches were used.
	
	Transmission and reflection spectra of finite Ta$_2$O$_5$-coated PS lattices were computed with a commercial FDTD solver (Lumerical FDTD).\cite{Ansys} The geometry consisted of 94 hexagonally packed PS spheres on a SiO$_2$ substrate, coated with a Ta$_2$O$_5$ layer of thickness $t_\mathrm{shell}$. PS spheres were modeled with refractive index $n = 1.59$, Ta$_2$O$_5$ with $n \approx 2.0$, and the glass substrate with $n = 1.45$. Normal-incidence plane-wave excitation and perfectly matched layers boundaries were used.
	
	Local-emitter quantities were obtained with a periodic-cell time-domain Maxwell solver (Tidy3D).\cite{FlexcomputeTidy3D} A single unit cell of the hexagonal lattice on a semi-infinite SiO$_2$ substrate was modeled, with a conformal Ta$_2$O$_5$ shell of thickness $t_\mathrm{shell}$ surrounding the PS sphere upper halves, periodic lateral boundary conditions and perfectly matched layers along the surface normal.
	
	Classical electric dipoles were placed either on Ta$_2$O$_5$ caps or in the valleys between neighbor caps, with in-plane ($E_x$) and out-of-plane ($E_z$) dipole orientations. The total radiated power and the power crossing a plane above the metasurface were compared to a reference dipole in a homogeneous host to extract $F_{\mathrm{p}}(\lambda)$ and $\beta_{\mathrm{top}}(\lambda)$ via Eqs.~\eqref{eq:Purcell} and \eqref{eq:beta}. Emission-weighted metrics $F_{\mathrm{p}}^{\mathrm{em}}$ and $F_{\mathrm{p,rad}}^{\mathrm{em}}$ were obtained by weighting with the measured Rh6G emission spectrum [Eqs.~\eqref{eq:Fpem} and \eqref{eq:Fpemrad}].

  Details of the emitter-placement averaging model used for comparison with ensemble Purcell-factor measurements are given in the Supporting Information (Sections S2-S8).
	
	\section{Results and discussion}
	
	\subsection{Thickness-tunable resonances in Ta$_2$O$_5$-coated lattices}
	
\begin{figure}[t]
		\centering
		% replace by your actual experimental RT figure
		\includegraphics[width=0.9\linewidth]{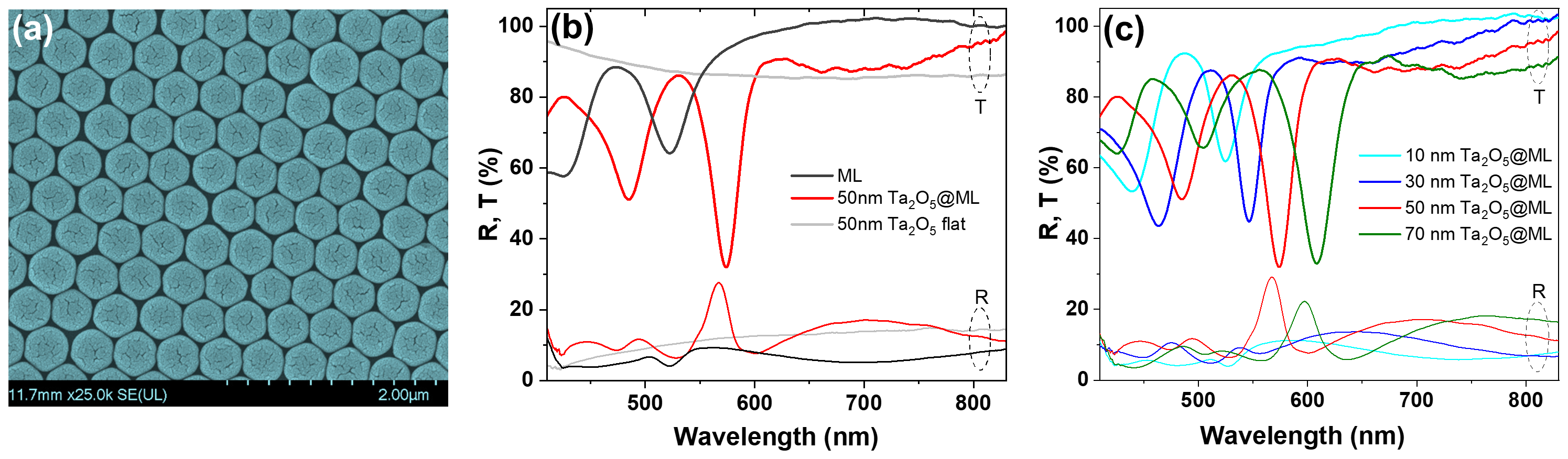}
		\caption{
        (a) SEM image of the two-dimensional PS microsphere monolayer. 
        (b and c) Experimental transmission and reflectance of Ta$_2$O$_5$-coated PS microsphere lattices. 
			(b) Comparison between a PS monolayer, the same monolayer coated with a \SI{50}{nm} Ta$_2$O$_5$ shell, and a flat \SI{50}{nm} Ta$_2$O$_5$ film on glass. 
			(c) Metasurfaces with Ta$_2$O$_5$ shell thicknesses of 10, 30, 50 and \SI{70}{nm}, showing a systematic red-shift of the resonance with increasing thickness.}
		\label{fig:RT_expt}
	\end{figure}

	Figure~\ref{fig:RT_expt} shows the typical two-dimensional lattice structure of the PS microsphere monolayer, as it results from the well-controlled self-assembly process. The experimental transmission and reflection spectra for a bare PS microsphere monolayer, the same monolayer coated with a \SI{50}{nm} Ta$_2$O$_5$ shell, and a flat \SI{50}{nm} Ta$_2$O$_5$ film on glass are presented in Figure~\ref{fig:RT_expt}b-c. The PS monolayer alone exhibits a broad transmission minimum around $\sim\SI{520}{nm}$, characteristic of Mie-type resonances of the microspheres. After coating with Ta$_2$O$_5$, this minimum red-shifts by almost \SI{60}{nm}, and the corresponding reflectance peak becomes more pronounced, signaling the formation of hybrid PS/Ta$_2$O$_5$ photonic resonances. In contrast, the flat Ta$_2$O$_5$ film displays only smooth thin-film interference fringes without sharp features.

    The Ta$_2$O$_5$-coated PS lattices exhibit a pronounced transmission dip accompanied by a reflectance peak, whose wavelength red-shifts with $t_\mathrm{shell}$. This behavior is consistent with a resonance condition defined by Eq.~\eqref{eq:resonance}
	\begin{equation}
		m\,\lambda \simeq 2\,n_\mathrm{eff}\,d,
		\label{eq:resonance}
	\end{equation}
	where $m$ is an integer order, $n_\mathrm{eff}$ an effective refractive index, and $d$ an effective thickness of the hybrid shell--lattice layer. Increasing $t_\mathrm{shell}$ increases $n_\mathrm{eff}d$, shifting the resonance to a longer wavelength. The same figure also presents spectra for Ta$_2$O$_5$-coated PS lattices with $t_\mathrm{shell} = \SI{10}{nm}, \SI{30}{nm}, \SI{50}{nm}$, and \SI{70}{nm}. The transmittance minima occur near 526, 547, 576, and \SI{608}{nm}, respectively, demonstrating a clear and nearly monotonic red-shift with increasing shell thickness. This trend follows the simple effective-thickness picture of Eq.~\eqref{eq:resonance}, and confirms that the Ta$_2$O$_5$ shell provides a convenient handle to tune the lattice resonance across the visible range.
	
	Finite-cluster FDTD simulations of a Ta$_2$O$_5$ shell on a 94-sphere PS lattice reproduce these trends (Fig.~\ref{fig:FDTD_cluster}). The simulated transmission dip and reflection peak for each shell thickness closely match the experimental positions and linewidths (Fig.~S2), with residual differences attributable to fabrication tolerances and finite-size effects. The agreement supports the picture that the experimental Ta$_2$O$_5$ growth is effectively conformal at the scale relevant for the optical response.
	
	\subsection{Fluorescence enhancement and resonance alignment}
	
\begin{figure}[t]
		\centering
		% replace by your actual PL figure
		\includegraphics[width=0.9\linewidth]{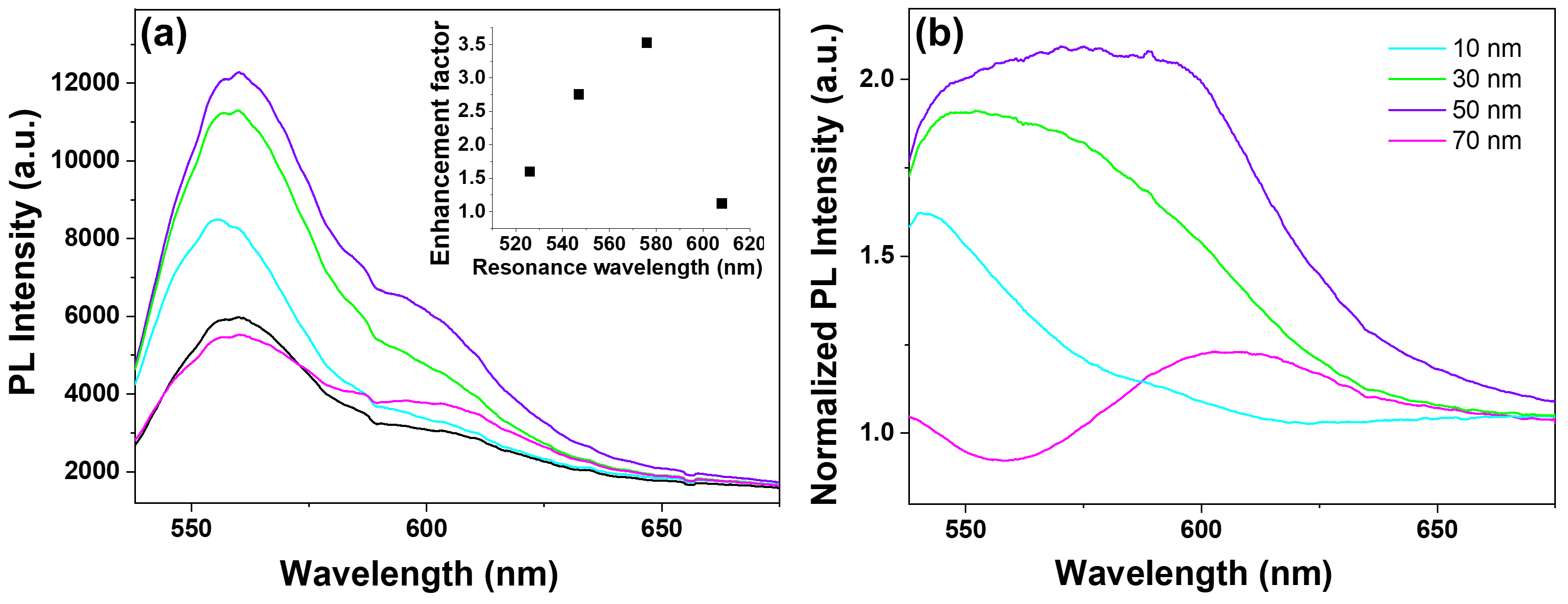}
		\caption{Fluorescence of Rh6G/PVP on Ta$_2$O$_5$-coated PS microsphere lattices. 
			(a) Emission spectra for metasurfaces with Ta$_2$O$_5$ shells of 10, 30, 50 and \SI{70}{nm} and for the glass reference (black line). 
			(b) Normalized enhancement spectra (sample/reference), highlighting spectral regions of maximum enhancement and their correlation with the lattice resonances. 
			Inset: integrated fluorescence enhancement factor $E_{\mathrm{F}}$ as a function of resonance wavelength for the different shell thicknesses.}
		\label{fig:PL}
	\end{figure}

	Figure~\ref{fig:PL} presents representative Rh6G/PVP fluorescence spectra for the reference on glass and for Ta$_2$O$_5$-coated PS lattices with $t_\mathrm{shell} = \SI{10}{nm}, \SI{30}{nm}, \SI{50}{nm}$ and \SI{70}{nm}. All metasurfaces exhibit enhanced emission intensity relative to the glass reference. The enhancement is particularly strong for the \SI{30}{nm} and \SI{50}{nm} shells, consistent with their resonances falling within the Rh6G emission band and, for the thinner sample, near the excitation wavelength.
	
	To isolate spectral shaping effects, the emission from each metasurface is divided by the reference spectrum, yielding normalized enhancement spectra. For each shell thickness, the region of maximum enhancement coincides with the transmission minimum of the corresponding lattice resonance in Fig.~\ref{fig:RT_expt}. The \SI{10}{nm} shell shows enhancement mainly around $\sim\SI{540}{nm}$, the \SI{30}{nm} shell around $\sim\SI{550}{nm}$, the \SI{50}{nm} shell exhibits broad enhancement across most of the Rh6G band and centered near $\sim\SI{570}{nm}$, and the \SI{70}{nm} shell displays enhancement predominantly around $\sim\SI{605}{nm}$. Micro-spectroscopy measurements reveal some spatial heterogeneity for the thickest shells and hints of secondary emission bands associated with higher-order photonic modes, but the global trends remain robust.
    
    The fluorescence enhancement factor for each Ta$_2$O$_5$ metasurface was quantified owing to the integration performed over the measured Rh6G emission spectrum according to Eq.~\eqref{eq:Fpemrad}. The inset of Fig.~\ref{fig:PL} summarizes the integrated enhancement factor as a function of the measured resonance wavelength for each Ta$_2$O$_5$ thickness. The enhancement does not simply grow with increasing shell thickness. Rather, it reflects a balance between (i) excitation-field enhancement when the resonance lies close to the pump wavelength and (ii) LDOS and out-coupling enhancement when the resonance overlaps the emission band,\cite{Lakowicz,LodahlNature2004} as well as the spacer role of Ta$_2$O$_5$ in mitigating quenching at very small separations. The \SIrange{30}{50}{nm} shells achieve the most favorable compromise, yielding the largest integrated enhancements across the measurement conditions. Increasing the Ta$_2$O$_5$ shell thickness leads to a progressive red-shift of the optical resonance further away from the \SI{532}{nm} excitation wavelength and beyond the emission maximum of Rh6G. The fluorophore molecules experience weaker local electric field enhancement at the excitation wavelength and fewer fluorophore molecules are excited. As it can also be seen in the normalized spectra in Fig.~\ref{fig:PL}, the \SI{70}{nm} shell enhances mainly the red edge of Rh6G emission around \SIrange{600}{610}{nm}, leading to reduced integrated intensity.
	
\begin{figure}[t]
		\centering
		% replace by your actual PL figure
		\includegraphics[width=0.9\linewidth]{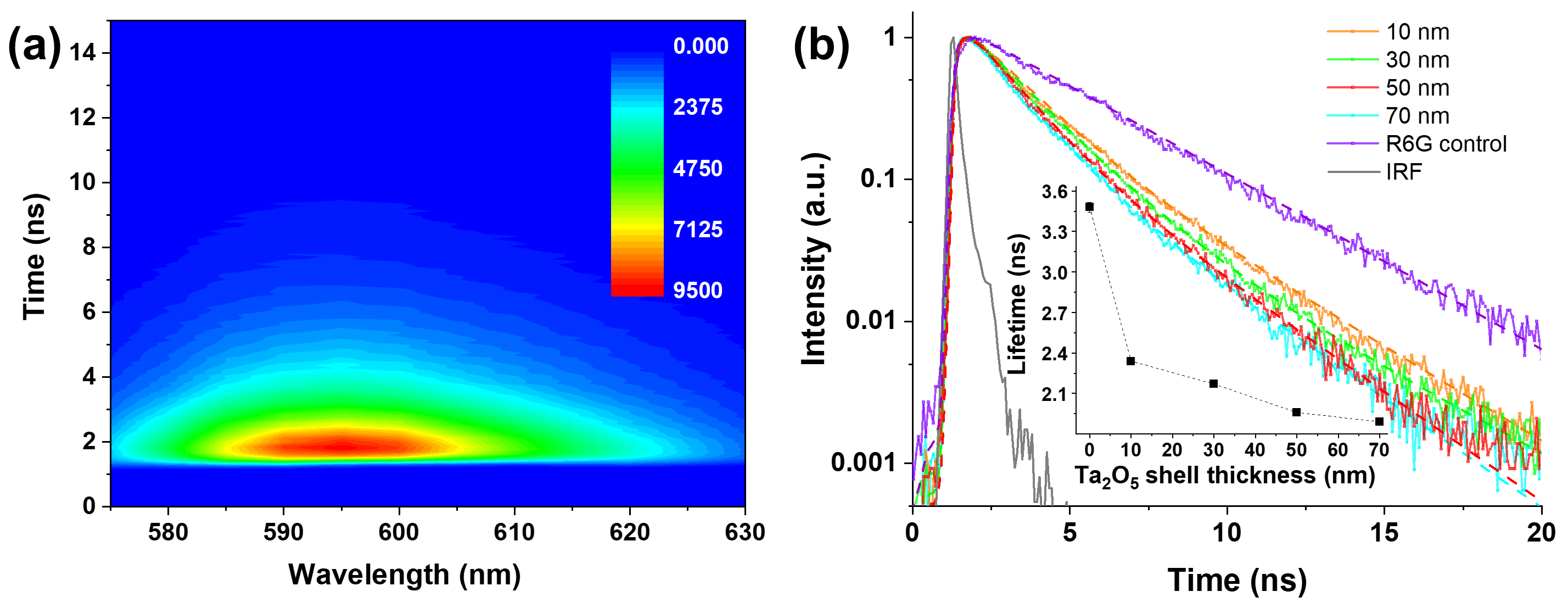}
		\caption{
        (a) Time-correlated single photon counting (TCSPC) data showing the three-dimensional data carpet obtained from a 50 nm shell thickness Ta$_2$O$_5$ metasurface when pumped at 545 nm. 
        (b) Normalized kinetic traces recorded at the 590 nm emission maximum. The dotted lines represent the corresponding fitted curves. The inset shows the calculated lifetimes characteristic to the Rhodamine 6G control sample and to each metasurface with Ta$_2$O$_5$ shells of 10, 30, 50 and 70 nm.}
		\label{fig:TCSPC}
	\end{figure}

    Figure ~\ref{fig:TCSPC} presents the TCSPC decay traces recorded at the emission maximum of 590 nm for Ta$_2$O$_5$ coated microsphere lattices and the control Rh6G/PVP film deposited on glass, respectively. The experimental decays are shown together with the reconvolution fits obtained using the measured IRF. The decay of the fluorophore on Ta$_2$O$_5$ metasurfaces results in a systematic reduction of the  decay time, whereas fluorophores on the control sample exhibit the longest fluorescence lifetime.
    The average intensity-weighted lifetimes, summarized in Fig.~\ref{fig:TCSPC}b, decrease monotonically with increasing Ta$_2$O$_5$ shell thickness, from 3.48 ns for the control sample to 1.89 ns for the 70 nm coating. This progressive lifetime shortening indicates a thickness-dependent increase in the total decay rate of Rh6G, consistent with the modification of the local photonic environment and the enhanced coupling of molecular emission to photonic modes supported by the high-index Ta$_2$O$_5$ metasurface.
    The steady-state fluorescence enhancement and time-resolved measurements provide complementary information on the light–matter interaction in the Ta$_2$O$_5$ metasurfaces. Although the integrated fluorescence enhancement reaches a maximum for intermediate shell thicknesses (30–50 nm), the fluorescence lifetime decreases monotonically with increasing Ta$_2$O$_5$ thickness. This indicates that although thicker coatings enhance the total decay rate through increased local density of optical states, maximal detected fluorescence occurs at thicknesses where excitation-field enhancement and radiative out-coupling are optimally balanced. Thus, fluorescence enhancement in the present system is governed by the interplay between excitation enhancement, LDOS modification, and emission extraction efficiency rather than by decay-rate acceleration alone.

	\subsection{Finite-cluster simulation of the lattice resonances}
	
\begin{figure}[t]
		\centering
		% replace by your actual FDTD cluster figure (old Fig. 5)
		\includegraphics[width=0.9\linewidth]{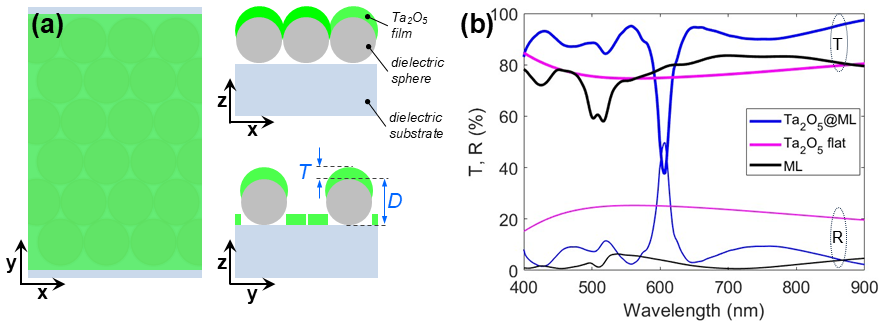}
		\caption{Finite-cluster FDTD simulations of Ta$_2$O$_5$-coated PS lattices. 
			(a) Schematic of the simulated structure: a hexagonal cluster of PS spheres on a glass substrate coated by a Ta$_2$O$_5$ shell and residual film. 
			(b) Simulated transmittance and reflectance spectra for a flat Ta$_2$O$_5$ film, a PS microsphere monolayer, and a Ta$_2$O$_5$-coated microsphere monolayer. The hybrid lattice exhibits a deeper and red-shifted resonance compared to the bare PS monolayer, in agreement with experiment.}
		\label{fig:FDTD_cluster}
	\end{figure}

	To connect the experimental resonances to the underlying modal structure, we performed FDTD simulations of a finite-size Ta$_2$O$_5$-coated PS cluster. Figure~\ref{fig:FDTD_cluster}(a) schematically illustrates the simulated geometry: a hexagonal patch of PS spheres on glass, covered by a conformal Ta$_2$O$_5$ shell and residual film. Figure~\ref{fig:FDTD_cluster}(b) compares the simulated transmission and reflection spectra of (i) a flat Ta$_2$O$_5$ film, (ii) a bare PS monolayer, and (iii) a Ta$_2$O$_5$-coated PS monolayer for a representative shell thickness (e.g.\ \SI{70}{nm}).
	
	The bare PS monolayer supports a modest transmission dip and a reflection peak near $\sim\SI{520}{nm}$. Adding a Ta$_2$O$_5$ shell yields a deeper transmission minimum and stronger reflection maximum red-shifted toward 540--\SI{550}{nm}, consistent with the experimental red-shifts upon coating (Fig.~\ref{fig:RT_expt}). The flat Ta$_2$O$_5$ film lacks sharp resonances, showing only smooth interference fringes. Simulations of hypothetical hollow Ta$_2$O$_5$ semishells, obtained by removing the PS cores, reveal related but spectrally shifted resonances (Fig.~S3). These results confirm that the sharp spectral features observed experimentally originate from the hybrid PS/Ta$_2$O$_5$, with both core and shell jointly determining the effective index and mode structure.
	
	Electric-field maps at and away from resonance (Fig.~S3) show that, at the resonance wavelength, the fields are strongly enhanced in the upper portion of the Ta$_2$O$_5$ shell and in the interstitial regions, precisely where the Rh6G/PVP layer resides. This spatial overlap provides the basis for a natural framework for interpreting the fluorescence enhancement in terms of LDOS and directional out-coupling.
	
	\subsection{Single-emitter metrics and ensemble Purcell factors}
	
	The periodic-cell simulations yield the wavelength-dependent Purcell factor $F_{\mathrm{p}}(\lambda)$ and the product $F_{\mathrm{p}}(\lambda)\,\beta_{\mathrm{top}}(\lambda)$ for single dipoles at well-defined locations and orientations. Figure~S4 in the Supporting Information summarizes these results for shell thicknesses $t_\mathrm{shell} = \SI{10}{nm}, \SI{30}{nm}, \SI{50}{nm}, \SI{70}{nm}$, with panels distinguishing emitters on caps vs between caps and in-plane vs out-of-plane dipole orientations.
	
	Several trends emerge. First, for all thicknesses the averaged ($F_{\mathrm{p}}^{\mathrm{em}} = <F_p>$, Eq. \eqref{eq:Fpem}) in-plane dipoles generally experience larger $F_{\mathrm{p}}$ than out-of-plane ones, especially when located on cap of the spheres, but also between caps for larger thicknesses of the coating layer. This reflects the predominantly tangential character of the hybrid lattice modes. Second, the product $F_{\mathrm{p}}\,\beta_{\mathrm{top}}$, which controls the contribution to top-side emission [Eq.~\eqref{eq:Idet}], is maximized for in-plane dipoles located on or near the upper Ta$_2$O$_5$ caps for intermediate thicknesses (\SIrange{30}{50}{nm}). Third, the spectral position of the $F_{\mathrm{p}}\,\beta_{\mathrm{top}}$ maximum shifts to longer wavelengths as $t_\mathrm{shell}$ increases, tracking the resonance movement in Fig.~\ref{fig:RT_expt}.
	
	When these single-emitter responses are integrated over the Rh6G emission spectrum to form $F_{\mathrm{p,rad}}^{\mathrm{em}}$ [Eq.~\eqref{eq:Fpemrad}], the resulting values show the same qualitative thickness dependence as the measured fluorescence enhancement $E_{\mathrm{F}}$ in Fig.~\ref{fig:PL}. In particular, the intermediate Ta$_2$O$_5$ thicknesses provide the best compromise between LDOS enhancement and efficient out-coupling into the upper half-space for emitters located near the top of the shells. This picture is consistent with the experimental observation that the \SIrange{30}{50}{nm} shells yield the largest integrated enhancement.
	
	\subsection{Purcell factor for PS/Ta$_2$O$_5$ microsphere lattices: theory vs experiment}
	
	To establish an emitter--environment model and quantify how single-dipole responses translate into ensemble measurements, we consider the PS/Ta$_2$O$_5$ microsphere lattice with the same Ta$_2$O$_5$ shell thicknesses as studied above. For this geometry, the experimental observable---the Purcell factor extracted from fluorescence lifetime measurements---averages over a large number of emitter configurations, making it an ideal benchmark for an averaging framework.
	
	We model a PS sphere of radius $R_\mathrm{sphere} = \SI{230}{nm}$ (nominal diameter $D = \SI{460}{nm}$) coated with a Ta$_2$O$_5$ shell of thickness $t_\mathrm{shell} \in \{10,30,50,70\}\,\mathrm{nm}$. Dye emitters are assumed to be distributed on the outer shell surface, either on the spherical caps (high-curvature regions) or in the valleys between caps, as motivated by the wetting and drying dynamics during spin-coating. For each placement, we compute $F_{\mathrm{p}}(\lambda)$ for in-plane and out-of-plane dipole orientations, and then construct an averaged Purcell factor $\langle F_{\mathrm{P}}(\lambda)\rangle$ that accounts for (i) the probability of being on a cap vs between caps, (ii) random dipole orientation, and (iii) ensemble spectral smoothing. Details of this averaging model are given in Sections~S3--S7 of the Supporting Information.
	
	In compact form, the spatially and polarization-averaged Purcell factor reads (Eq.~\eqref{eq:FPavg_main}):
	\begin{equation}
		\langle F_{\mathrm{P}}(\lambda) \rangle = P_\mathrm{on}\,\langle F_{\mathrm{P}}^{\mathrm{on}}(\lambda) \rangle_{\mathrm{pol}}
		+ P_\mathrm{bet}\,\langle F_{\mathrm{P}}^{\mathrm{bet}}(\lambda) \rangle_{\mathrm{pol}},
		\label{eq:FPavg_main}
	\end{equation}
	where $P_\mathrm{on}$ and $P_\mathrm{bet}=1-P_\mathrm{on}$ denote the probabilities for an emitter to reside on a cap or between caps, respectively, and the polarization averages follow the standard isotropic weighting (two transverse and one longitudinal component). The probabilities are modeled as a logistic function of the
	dimensionless curvature ratio $t_\mathrm{shell}/R_\mathrm{sphere}$, derived from a two-state Boltzmann partition between capillary suction into the interstices and contact-line pinning on the caps \cite{deGennes2004, Kramers1940}, and constrained by a global optimization across all four shell thicknesses (see SI).
	
	In addition, we introduce two per-thickness corrections when comparing to experimental data: (i) a spectral smoothing factor $s \in [0,1]$ that damps oscillations present in single-geometry simulations but absent in ensemble measurements, and (ii) a small wavelength shift $\Delta\lambda$ that accounts for modest discrepancies between nominal and actual shell thicknesses and refractive indices. The full model is given by Eq.~\eqref{eq:FPmodel_main}
	\begin{equation}
		F_{\mathrm{P}}^{\mathrm{model}}(\lambda) = \overline{F}_{\mathrm{P}} 
		+ s \big[ \langle F_{\mathrm{P}}(\lambda + \Delta\lambda)\rangle - \overline{F}_{\mathrm{P}} \big],
		\label{eq:FPmodel_main}
	\end{equation}
	where $\overline{F}_{\mathrm{P}}$ is the spectral average of $\langle F_{\mathrm{P}}\rangle$ over the measured wavelength range. Global parameters controlling the curvature dependence are shared across all thicknesses, while $s$ and $\Delta\lambda$ are adjusted per shell.
	
\begin{figure}[t]
		\centering
		% replace by your actual Purcell comparison figure (old Fig. 7)
		\includegraphics[width=0.9\linewidth]{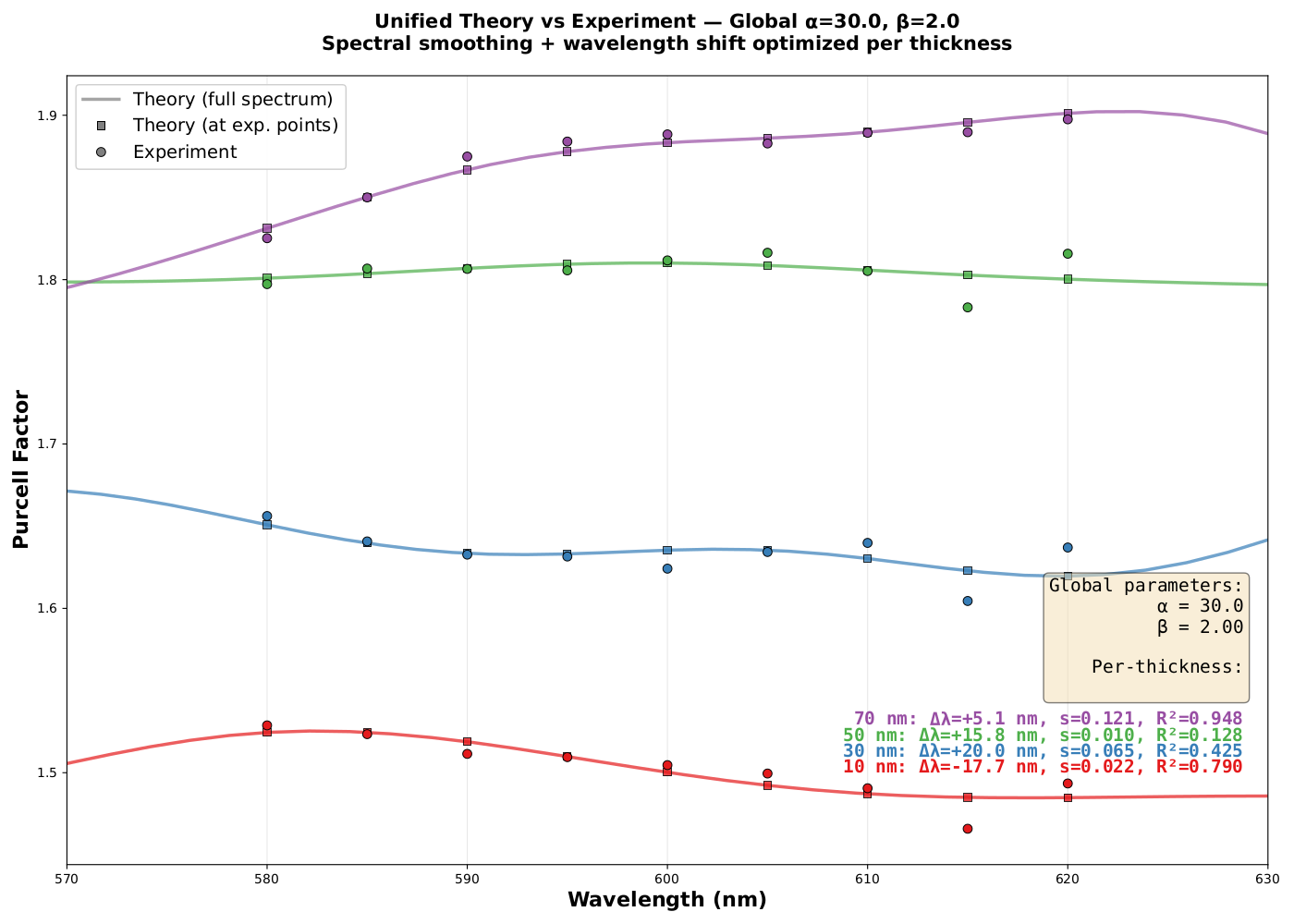}
		\caption{Comparison between the modeled Purcell factor (solid lines) and experimental data (circles) for PS/Ta$_2$O$_5$ microsphere lattices with Ta$_2$O$_5$ shell thicknesses of 10, 30, 50 and \SI{70}{nm} (bottom to top). Squares indicate the model evaluated at the experimental wavelengths. Global curvature parameters are shared across all thicknesses; per-thickness smoothing factors $s$ and wavelength shifts $\Delta\lambda$ are indicated in the inset.}
		\label{fig:Purcell_lattice}
	\end{figure}

	Figure~\ref{fig:Purcell_lattice} compares the resulting $F_{\mathrm{P}}^{\mathrm{model}}(\lambda)$ to the experimentally extracted Purcell factors for PS/Ta$_2$O$_5$ microsphere lattices with $t_\mathrm{shell} = 10, 30, 50, 70\,\mathrm{nm}$. The agreement is very good for the \SI{70}{nm} and \SI{10}{nm} shells, with coefficients of determination $R^2$ close to 0.95 and 0.8, respectively, and remains qualitatively correct for intermediate thicknesses. The optimized curvature parameters yield a monotonic increase of the on-cap probability from $\sim 0.05$ at \SI{10}{nm} to $\sim 0.74$ at \SI{70}{nm}, consistent with the progressive filling of the interstices
	and the resulting loss of capillary suction as the shell thickness grows (see SI for details). The small smoothing factors $s < 0.13$ indicate that the experimental spectra effectively average over a very large number of geometric configurations, far exceeding the discrete set sampled in the simulations.
	
	This agreement shows that the same physical ingredients that control emitter--lattice coupling in the Ta$_2$O$_5$-coated PS metasurfaces---namely, the spatial distribution of emitters relative to high-LDOS regions, dipole orientation averaging, and ensemble spectral smoothing---govern the Purcell factors measured on the microsphere lattices. The single-emitter responses summarized in Fig.~S4 thus provide a consistent foundation for interpreting both types of experiments.
	
	\section{Conclusions}
	
	We have presented a combined experimental and numerical study that establishes Ta$_2$O$_5$-coated PS microsphere lattices as dielectric SEF plastforms. The central finding is that a single geometric parameter, the Ta$_2$O$_5$ shell thickness, provides continuous and predictable control over the lattice photonic resonance across approximately \SI{80}{nm} of the visible spectrum, from \SIrange{526}{608}{nm} for shells of \SIrange{10}{70}{nm}, without lithographic patterning. The result is a scalable route to resonance-emitter overlap optimization that is not available in fixed-geometry plasmonic or lithographic alternatives.
    
   The framework developed to interpret these observations operates at three connected scales. Finite-cluster FDTD simulations reproduce the experimental far-field spectra and implicitly confirm that the deposited Ta$_2$O$_5$ grows conformally on the sphere lattice, a non-trivial structural validation. Periodic-cell dipole simulations yield wavelength-dependent Purcell factors and directional $\beta$-factors, allowing excitation enhancement, LDOS modification and emission out-coupling to be disentangled. The observation that the integrated fluorescence enhancement peaks at intermediate shell thicknesses (\SIrange{30}{50}{nm}), while the fluorescence lifetime decreases monotonically across all thicknesses, is a direct experimental manifestation of this three-way interplay. Our results show that fluorescence enhancement is governed by three competing factors, i.e. excitation enhancement, LDOS modification, and out-coupling efficiency, all controllable by shell morphology. A physically grounded averaging model based on capillary wetting arguments connects single-emitter simulations to measured Purcell factor spectra with shared global parameters across all four shell thicknesses, providing a stringent validation of the physical picture.

    The significance of this framework lies in its transferability. The same chain of geometric modeling, LDOS simulation and ensemble averaging can be applied prospectively — given a target emitter band, the optimal shell thickness, and emitter placement can be identified computationally before fabrication. The approach generalizes naturally to other high-index dielectrics compatible with conformal deposition (TiO$_2$, HfO$_2$, Nb$_2$O$_5$) and to other emitter systems, including quantum dots and J-aggregates, where the absence of metallic quenching is especially valuable. The Ta$_2$O$_5$ geometry naturally integrates with planar Ta$_2$O$_5$ photonics,\cite{Liu2025,Brodnik2025} opening a path toward hybrid waveguide-metasurface configurations for evanescent-field fluorescence sensing. More broadly, our results provide a quantitative foundation for rational design of colloidal-based dielectric metasurfaces as a class of systems bearing fabrication simplicity and material versatility.

   \section*{Supporting Information} The material supplied as Supporting Information describes additional UV-visible simulations for Ta$_2$O$_5$-coated PS lattices, showing the role of PS microsphere diameter and Ta$_2$O$_5$ shell thickness, the finite-cluster FDTD simulations exhibiting the field distirbutions in Ta$_2$O$_5$-PS lattices, as well as a single-emitter analysis for the Ta$_2$O$_5$ metasurfaces used as building blocks for the Purcell-factor model presented in the main text together with a statistical distribution of emitter positions that enables the comparison of the single-emitter simulations with the experimentally measured Purcell factors. 
	
	\section*{Acknowledgements}
	
	This work was supported by a grant of the Romanian Ministry of Research, Innovation and Digitalization (MCID), CCDI--UEFISCI, project number PN-IV-P8-8.3-PM-RO-FR-2024-0163, within PNCDI IV. Authors from INCDTIM also acknowledge the ``Nucleu'' Program within the National Plan for Research, Development and Innovation 2022--2027, project PN 23 24 01 02. The authors thank I.~Cardan and D.~Cuibus for help with preparing microsphere arrays.

\newpage

\textbf{TOC graphic}

	\begin{figure}[h]
		\centering
		\includegraphics[width=1\linewidth]{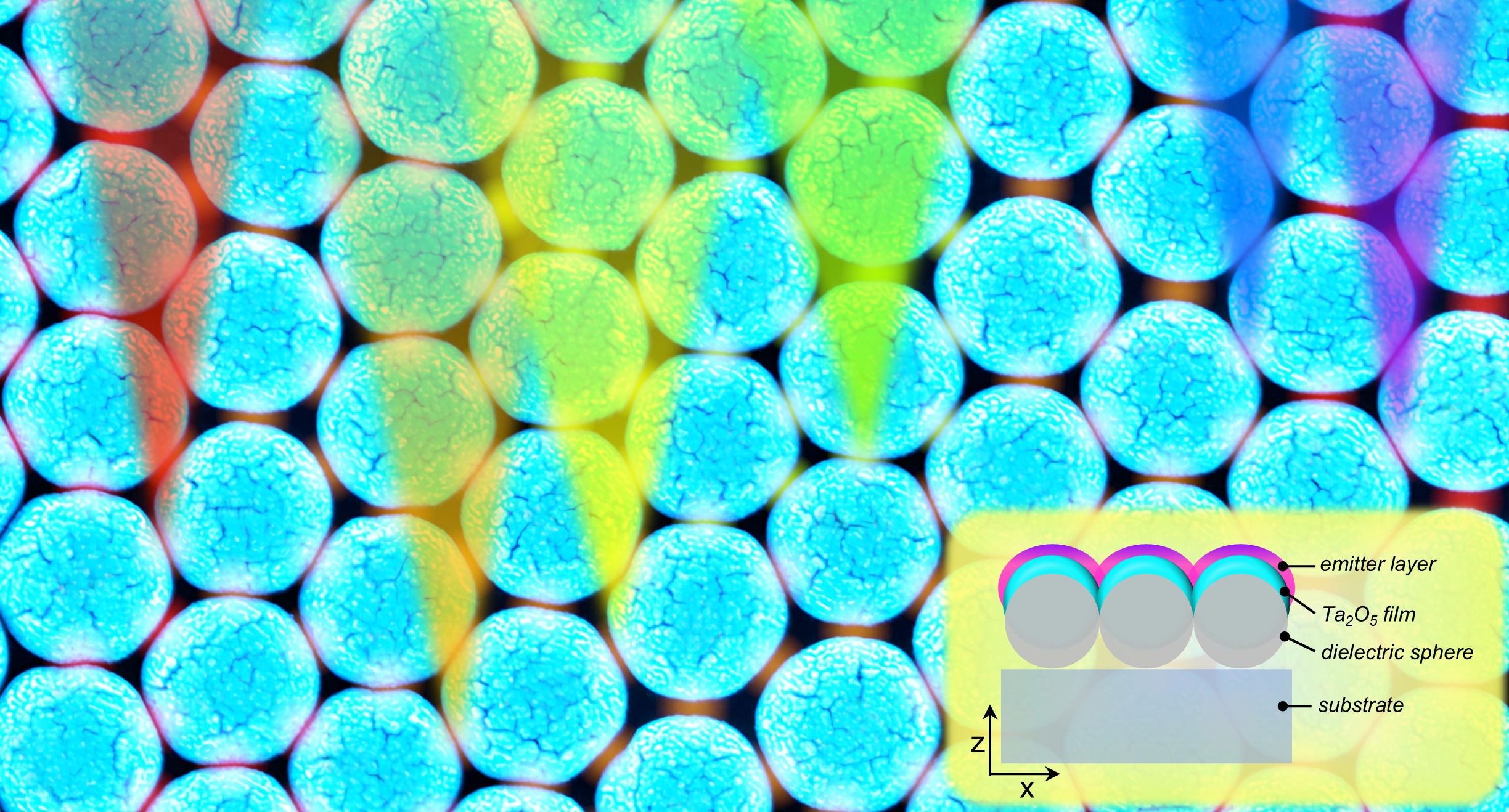}
	\end{figure}
	

\begin{thebibliography}{99}
	
		\bibitem{Kuznetsov2016}
		Kuznetsov, A.~I.; Miroshnichenko, A.~E.; Brongersma, M.~L.; Kivshar, Y.~S.; Luther-Davies, B., Optically resonant dielectric nanostructures, Science, \textbf{2016}, 354(6314), aag2472.
		
		\bibitem{Staude2017}
		Staude, I.; Schmid, J., Metamaterial-inspired silicon nanophotonics, Nat. Photonics, \textbf{2017}, 11, 274--284.
		
		\bibitem{Liu2016}
		Liu, W.; Kivshar, Y.~S., Multipolar interference effects in nanophotonics, Philos. Trans. A Math. Phys. Eng. Sci., \textbf{2017}, 375(2090), 20160317.
		
		\bibitem{NovotnyHecht}
		Novotny, L.; Hecht, B., \textit{Principles of Nano-Optics}, Cambridge University Press, \textbf{2012}.
		
		\bibitem{LodahlRMP}
		Lodahl, P.; Mahmoodian, S.; Stobbe, S., Interfacing single photons and single quantum dots with photonic nanostructures, Rev. Mod. Phys., \textbf{2015}, 87(2), 347--400.
		
		\bibitem{Lakowicz}
		Lakowicz, J.~R., \textit{Principles of Fluorescence Spectroscopy}, Springer, Boston, 3rd edn, \textbf{2006}.
		
		\bibitem{Kuehn2006}
		K\"{u}hn, S.; Hakanson, U.; Rogobete, L.; Sandoghdar, V., Enhancement of Single-Molecule Fluorescence Using a Gold Nanoparticle as an Optical Nanoantenna, Phys. Rev. Lett., \textbf{2006}, 97, 017402.
		
		\bibitem{LodahlNature2004}
		Lodahl, P.; van Driel, A.~F.; Nikolaev, I.~S.; Irman, A.; Overgaag, K.; Vanmaekelbergh, D.; Vos, W.~L., Controlling the dynamics of spontaneous emission from quantum dots by photonic crystals, Nature, \textbf{2004}, 430, 654--657.
	
		\bibitem{Mie1908}
		Mie, G., Beitr\"{a}ge zur Optik tr\"{u}ber Medien, speziell kolloidaler Metall\"{o}sungen, Annalen der Phys., \textbf{1908}, 330(3), 377--445.

        \bibitem{Nechita}
        Nechita, N.; Farcau, C.; Tuning plasmons of metal-coated microsphere arrays towards optimized surface-enhanced spectroscopy, Optics Express \textbf{2021}, 29(25), 42238-42250.
        	
		\bibitem{Steinlechner2018}
		Steinlechner, J., Development of mirror coatings for gravitational-wave detectors, Philos. Trans. R. Soc. A, \textbf{2018}, 376, 20170282.
		
		\bibitem{Belt2017}
		Belt, M.; Davenport, M.~L.; Bowers, J.~E.; Blumenthal, D.~J., Ultra-Low-Loss Ta$_2$O$_5$-core/SiO$_2$-clad planar waveguides on Si substrates, Optica, \textbf{2017}, 4(5), 532--535.
		
		\bibitem{Wang2021}
		Wang, T.-J.; Chen, P.-K.; Li, Y.-T.; Sung, A.-N., Athermal high-Q tantalum-pentoxide-based microresonators on silicon substrates, Opt. Laser Technol., \textbf{2021}, 138, 106925.
		
		\bibitem{Subramanian2009}
		Subramanian, A.~Z.; Oton, C.~J.; Wilkinson, J.~S.; Greef, R., Waveguiding and photoluminescence in Er$^{3+}$-doped Ta$_2$O$_5$ planar waveguides, J. Luminesc., \textbf{2009}, 129(8), 812--816.

		\bibitem{ZhangLSA}
		Zhang, C.; Chen, L.; Lin, Z.; Song, J.; Wang, D.; Li, M.; Koksal, O.; Wang, Z.; Spektor, G.; Carlson, D.; Lezec, H.~J.; Zhu, W.; Papp, S.; Agrawal, A., Tantalum pentoxide: a new material platform for high-performance dielectric metasurface optics in the ultraviolet and visible region, Light Sci. Appl., \textbf{2024}, 13, 23.

		\bibitem{Schmitt2008}
		Schmitt, K.; Oehse, K.; Sulz, G.; Hoffmann, C., Evanescent field sensors based on tantalum pentoxide waveguides -- A review, Sensors, \textbf{2008}, 8(2), 711--726.
		
		\bibitem{Grandin2006}
		Grandin, H.~M.; St\"{a}dler, B.; Textor, M.; V\"{o}r\"{o}s, J., Waveguide excitation fluorescence microscopy: a new tool for sensing and imaging the biointerface, Biosens. Bioelectron., \textbf{2006}, 21(8), 1476--1482.

		\bibitem{Zhang}
		Zhang, J-T.; Wang, L.; Lamont, D. N; Velankar, S. S.; Asher, Sanford A., Fabrication of large-area two-dimensional colloidal crystals, Angew. Chem. Int. Ed., \textbf{2012}, 51(25), 6117--6120.

		\bibitem{Ansys}
		Ansys Lumerical FDTD Simulation of Photonic Components, \texttt{https://www.ansys.com/products/optics/fdtd}.

		\bibitem{FlexcomputeTidy3D}
		Flexcompute Inc., Tidy3D Electromagnetic Solver, \texttt{https://www.flexcompute.com/tidy3d/}.

		\bibitem{deGennes2004}
		P.-G.~de~Gennes, F.~Brochard-Wyart, and D.~Qu\'er\'e ,
Capillarity and Wetting Phenomena: Drops, Bubbles, Pearls, Waves,
		Springer, New York (2004).
		
		\bibitem{Kramers1940}
		Kramers, H. A., Brownian motion in a field of force and the diffusion model of chemical reactions, Physica, \textbf{1940}, 7, 284--304.
		
		\bibitem{Liu2025}
		Liu, Z.; Yao, W.; You, M.; Yu, X.; Ding, N.; Cheng, W.; Li, Z.; Tang, X.; Guo, F.; Zhao, Q., Tantalum Pentoxide Integrated Photonics: A Promising Platform for Low-Loss Planar Lightwave Circuits with Low Thermo-Optic Coefficients, ACS Photonics, \textbf{2025}, 12(3), 684--695.

        \bibitem{Brodnik2025}
		Brodnik, G.~M.; Spektor, G.; Williams, L.~M.; Zhang, J.; Carollo, A.~R.; Dan, A.; Black, J.~A.; Carlson, D.~R.; Papp, S.~B., Monolithic 3D integration of tantalum pentoxide photonics on arbitrary substrates, arXiv:2509.08092, \textbf{2025}.

	\end{thebibliography}
\end{document}